# Mediating Role of Managing Information Technology and Its Impact on Firm Performance: Insight from China

Abstract

Purpose – Managing IT with firm performance has always been a debatable topic in literature and practice. Prior studies examining the above relationship have reported mixed results and have yet ignored the eminent managing IT practices. The purpose of this paper is to empirically investigate the relevance of Val-IT 2.0 practice in managing IT investment, and its mediating role in the firm performance context.

Design/methodology/approach - This paper developed on two themes of literature. First managing IT as a firm's IT capability in order to generate value from IT investment. Second IT as a firm's resource under resource-based view offers firm's competence that deploys potentials in achieving firm performance. The structural equation modeling with PLS techniques used for analyzing data collected from 176 organization's IT, and business executives in China.

Findings - The results of this study show empirical evidence that Val-IT's components (value governance, portfolio management, and investment management) are significantly linked to the management of IT, and it found to be a significant mediator between Val-IT components and firm performance.

Research implications - This research contributes to the literature and practice by way of highlighting the value generation through managing IT on firm performance.

Originality - This study is fully based on Val-IT 2.0 with the firm performance where the managing IT mediate this relationship in a country-specific study in China. This study adds to the Chinese information system literature which suffers the lack of empirical studies in the context of management of IT research.

Paper type - Research paper

# 1. Introduction

For decades, executives and policy makers have always concerned the profitability of IT investment while it is increasing radically (Kim et al. 2009; Lee et al. 2011; Mithas et al. 2012; Prasad et al. 2010). It is obvious that IT investment can improve firm performance (Turel et al. 2017), but investments in IT are not sufficient by themselves to affect firm performance (Wang et al. 2015). Thus, it requires managing IT to realize its superior potential. The practice of managing IT, e.g., the managerial efforts related to planning, organizing, controlling, and directing the use of IT within an organization (Boynton and Zmud 1987; Van Der Zee and De Jong 1999; Wang et al. 2015) has received a considerable attention among information system (IS) scholars, and executives (Lowry and Wilson 2016; Mithas et al. 2012; Tallon et al. 2000; Xu et al. 2016).

China has been recognized that the world's manufacturing center (Mingzhi Li et al. 2015; Wang et al. 2016c) has transformed its economy to larger market orientation as a result, IT has been perceived to be an essential driver for their consequent business and economic success (Davison et al. 2008; Dologite et al. 1998). The participation in the global competition and the growing economy of China have called upon the massive accessibility of IT, and IT is believed to be an ever more critical resource (Chen 2010). Consequently, Chinese firms have heavily invested in IT infrastructure and various information systems in recent years (Peng et al. 2016; Shao et al. 2016). Further, the management of IT still ruins a new discipline and empirical research on IT issues in China are limited (Chen 2010; Wang et al. 2015).

The "IT productivity paradox" (inconsistency between massive IT investments and its lack of benefits) has been examined in many Chinese firms context (Kim et al. 2009; Peng et al. 2016). For example, studies aimed at identifying the real causes of IT productivity paradox, will likely to have significant implications for firms in China, as they struggle to generate business value from their IT investments (Peng et al. 2016). Besides, some prior studies in Chinese context are in industry-specific (Kim et al. 2009; Wang et al. 2016b), review based (Dologite et al. 1998; Li-Hua and Khalil 2006), and strategic planning oriented (Chi et al. 2017; Mao et al. 2016; Shao et al. 2016). Moreover, the prior researches on IS issues in China have targeted organizational, and system or software related issues and empirical research on the organization and strategic issues are scarce (Davison et al. 2008; Ji et al. 2007; Peng et al. 2016). Similarly, the Chinese IS management literature has focused the wide range of industrial, technical, and operational issues (Chen 2010; Davison et al. 2008). The management of IT or the multifaceted socio-technical study that influences on the firm performance has relatively received less attention. It is worth noting that, research on the effects of IT investment on firm performance in developing countries will be an essential area for future studies as well (Kim et al. 2009; Peng et al. 2016). Therefore, it is important for a theory-driven empirical research needs to be done to deepen how the management of IT can enrich the firm performance on IS literature in the Chinese firm context.

Considering the IT's rising significance to the firm's operations and performance, prior studies show the underlying mechanism of how IT investment is enhancing firm performance. For example, Prasad et al. (2010) found that IT governance initiatives are positively related with IT-related capabilities; subsequently it relates to internal process level performance, which in turn improves customer service and firm-level performance. Ali et al. (2015) presented the momentous link between IT investment governance components and corporate performance. Wu et al. (2015) studied the effect of IT governance mechanisms on organizational performance with strategic alignment as a mediator and found impactful linkage among them. However, the prior literature falls in short, to empirically examine the role of managing IT in firm performance, which subsequently needs a comprehensive framework covering a wide spectrum of IT management activities. Hence, this study integrates Val-IT 2.0 framework with the management of IT to examine how these effects on Chinese firm performance.

The Val-IT 2.0 framework is a comprehensive framework developed by IT Governance Institute (ITGI) that assimilates a set of practical governance principles, practices and supporting guidelines to help executives and enterprise leaders to optimize the realization of value from IT investments (Val 2008). Val-IT principles are applied in three domains, i.e., *VAlue goVErnance* - ensure value management practices secure optimal value from its IT-enabled investments during their full economic life cycle, *portfolio management* - ensure enterprise secures optimal value across its portfolio of IT-enabled investments, and *inVEstment management* - ensure enterprise's each IT-enabled investments contribute to value creation (Val 2008; Wilkin et al. 2012). Val-IT 2.0 is closely aligned and complements with COBIT from business and financial perspective, but delivers value to firms in its own right and supports professional who seeks IT investment decisions, and the realization of business value from IT (Lombardi et al. 2016; Val 2008; Wilkin et al. 2012). The extensive literature analysis of Lombardi et al. (2016) recognized that the Val-IT framework is the appropriate structure for managing IT investments and it combines all the main aspects of other IT governance models. Hence, Val-IT is suitable as a framework to study management of IT in a firm performance context. Based on the above discussion, we aim to address the following research questions:

RQ1. How are Val-IT components relevant in the management of IT investment context?

RQ2. How does the management of IT function as a mediator between Val-IT components and firm performance?

This study has several contributions to the literature and practices. The well-known Val-IT 2.0 framework for IT investment value delivery has either theoretically or empirically rarely been tested in the past. None of the prior studies used the actual Val-IT 2.0 framework both in global or country-specific research thus this study is the first attempt fully based on this framework which examines the management of IT with a firm performance for a country-specific study in China. Further, this study adds to the Chinese information system literature which suffers the lack of empirical studies in the context of managing IT research whereas technical and industry-wide studies are widely available.

The remaining of the paper is organized as follows. Section two discusses theoretical background with literature review. Section three proposes research model with hypothesis development. Section four presents research methodology and data analysis. Section five offers the findings of the results, and section six presents the discussion and implications. Finally, limitations with future research direction and conclusion are also given.

## 2. Theoretical Background

This section discusses two themes of literature that support this study. The first is managing IT as a firm's IT capability in order to generate value from IT investment. Second, the resource-based view (RBV) offers the firm's competence that deploys potentials in achieving firm performance.

### 2.1. IT capability and Value Delivery

As prior studies have examined the contributions of IS resources and capabilities to firm performance, the results are fragmented, and yet gaps exist in the literature (Ravichandran et al. 2005). Recently, researchers have studied that managing or governing IT is the firms' IT-related capabilities which leverage IT-enabled business value and in turn improve firm performance (Ali et al. 2015; Prasad et al. 2010; Turel et al. 2017; Wu et al. 2015; Zhang et al. 2016). Accordingly, a firm with a strong IT infrastructure and superior IT management ability could able to effectively deploy a new application, modify or redesign enterprise systems with structural sophistication, as well as solve maintenance hurdles (Chen et al. 2014; Wang et al. 2015). Further, Prasad et al. (2010) asserted that the prior studies found a number of IT-related capabilities in which IT-related management capabilities and IT-related infrastructure capabilities can be a basis of IT-related business value; thus firms with superior IT capability usually achieve superior firm performance (Zhang et al. 2016).

### 2.2. Resource-based view (RBV) of managing IT and firm performance

The theoretical insights of RBV provide a strong basis and demonstrate the importance of resource management with regard to firm performance (Mao et al. 2016; Xu et al.

2016). In IS literature, different authors have recognized various resources that are potential for firm performance such as human, technological, and relationship resources (Ravichandran et al. 2005); IT-related resources such as infrastructure, human-IT resources, and IT-enabled intangibles (Huang et al. 2006). More specific to IT, a firm competes and leverages from IT to create value because the resources it relies on are unique, rare, valuable, and costly to imitate under RBV (Prasad et al. 2010; Wang et al. 2016c). According to Xu et al. (2016), the more inimitable and diverse resources a firm possesses, the more possibility of competitive advantage it gains and sustains. Moreover, the RBV of a firm conceptualizes the potentials that firm realize from IT resources, can be the basis of its competitive advantage (Prasad et al. 2010; Turel et al. 2017; Zhang et al. 2016).

Drawing on the view of IT capability and RBV of the firm, managing IT is the firm's IT capability to create value from IT investment and IT as a firm's resource can be managed systematically to achieve firm performance. It is notable that the heterogeneous IT resources underpin by the resource-based view and IT capabilities make stronger the firm performance and preserve their impacts ahead in their future. According to Van Grembergen and De Haes (2009), IT is in a unique position to direct the business in adopting Val-IT practices, which in turn craft more value by levering IT to the firm. Hence, the inclusion of Val-IT practice with managing IT in this study is a panacea to exhibit underlying consequences between IT investment and firm performance.

## 3. Research Model and Hypotheses Development

The general rationale of this study is that Val - IT 2.0 components can facilitate for management of IT, which leads towards firm performance. In the research model, we propose that value governance, portfolio management, and investment management of Val-IT have a significant relationship with the management of IT. In addition, management of IT acts as a mediator between Val-IT 2.0 components and firm performance. We, therefore, develop our research model that not only based in theory, but also rooted inherently in practice that is suitable for examining Val-IT components and management of IT's impact on firm performance.

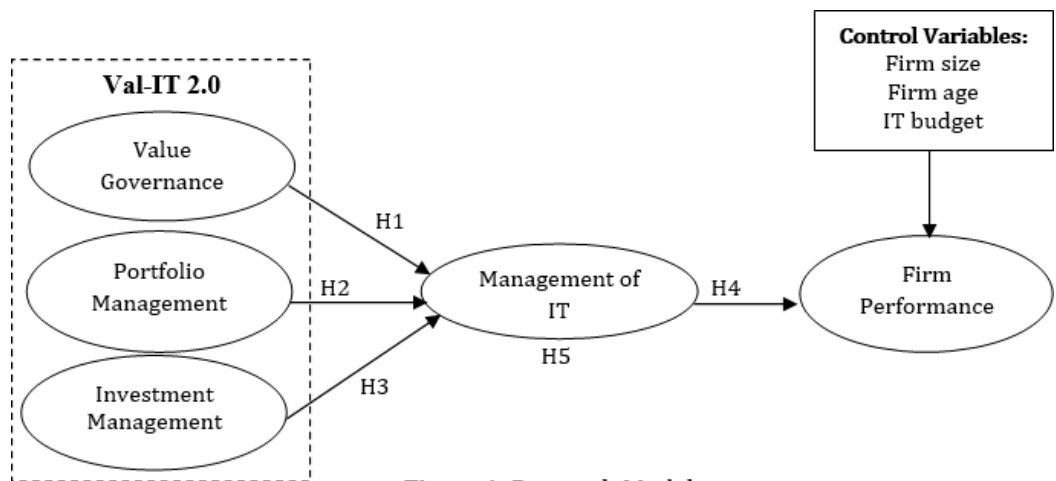

**Figure 1:** Research Model

Value Governance

Generating business value from IT has been an admired focal point in IS literature (Ali et al. 2015; Peng et al. 2016; Wilkin et al. 2016). The industry-wide surveys highlight the importance of value from IT investment. Accordingly, Gartner in 2002 found that 20 % of all expenditures on IT is wasted. An IBM survey in 2004 among Fortune 1000 CIOs shows that 40 % of all IT spending brought no return to their firms. Moreover, the

Standish Group study in 2006 found that 65 percent of IT projects were either challenged or failed (Val 2008). The value governance focuses on the structures and processes needed to ensure that value management practices are implanted in the firm and include ''necessary conditions" to enable a value-based approach in the consequent portfolio and investment management (Van Grembergen and De Haes 2009). In a typical firm, IT alone cannot directly create business value instead it creates through other firm resources and elements (Peng et al. 2016). For example, firm's both strategic alignment and IT investment evaluation are harmonizing to contribute higher IT business value and to realize how IT creates value for the firm (Tallon et al. 2000). Therefore, the firm's value governance is the management practice will likely have a reciprocal coalition with management of IT to leverage business value to improve firm performance. Thus, our first hypothesis is stated as follows.

H1: Firm with the greater ability of value governance practices will have a significant association with their management of IT.

Portfolio Management

Portfolio management focusses on the practices required to manage the whole portfolio of IT-enabled investments (Van Grembergen and De Haes 2009). Typically, a firm's effective management of portfolio of IT resources is a critical for IT executives (Chen 2012; Ferratt et al. 2005); as IT projects, mostly have greater uncertainty with a higher failure rate due to technological challenges such as hardware and software misconfiguration, network failure, security risks and interoperability issues (Wang et al. 2016a). Therefore, managing IT project is complicated and gives dilemmas to managers as it involves the conversion of business needs and workable solution while it holds many opportunities to the firm (Wang et al. 2016a). Given its growing prominence, the successful application of IT depends on the effective management of processes linked to the planning for, the acquisition of, and the execution of the firm's portfolio of IT (Boynton et al. 1994). Also, the proactive experience on IT projects and managing IT skills are critical to build IT competence for business managers (Bassellier et al. 2003). Further, the effective IT management avoids uncertainty in IT implementation and service delivery, and errors be corrected in earlier stages (Wang et al. 2015). Therefore, the second hypothesis is stated as follows.

H2: Firm with the greater ability of portfolio management practices will have a significant association with their management of IT.

Investment Management

IT investment is a firm's effort towards improvement, and generally, it is implemented by means of IT projects (Xu et al. 2016). IT investments can create business value with other enterprise resources and improve firm performance (Peng et al. 2016; Turel et al. 2017). While firms persistently seek effective managing IT strategies to better leverage IT investment (Lowry and Wilson 2016). The firm with a higher investment in IT and historical investment against occasional failures in their IT portfolio, and undergoes the virtuous cycle to become better at managing IT (Mithas et al. 2012). In recent year, substantial resources are consumed to manage IT, and the effective management of resources have been widespread for a long time (Prasad et al. 2010). This accords with the view that the top management decision making on effective IT management involves a synchronized effort in planning, organizing, monitoring, controlling, and directing to ensure that the expected value is delivered (Ali et al. 2015; Boynton and Zmud 1987; Prasad et al. 2010). Thus, our hypothesis is stated as follows.

H3: Firm with the greater ability of investment management practices will have a significant association with their management of IT.

Managing IT and Firm Performance

The impact of IT on firm performance is an evolving research area (Croteau and Bergeron 2001; Peng et al. 2016; Wu et al. 2015). The current managing IT is challengeable, that attains extensive attentions not only among executives and policy makers, but also investors and funding agencies in China (Li-Hua and Khalil 2006). The successful use of IT critically relies on managing IT and governance practices which are critically important to its value generation from IT investment (Ali et al. 2015; Prasad et al. 2010; Wu et al. 2015). Accordingly, aligning process-level benefits to firm-level outcomes is crucial, and hence considerable financial resources are devoted to obtaining and managing IT resources (Prasad et al. 2010). With the strong management of IT, firms could contribute to its performance by synchronizing activities across different business units, simplify operation processes, lower production cost, coordinate IT and business units, frequently check IT priorities, and timely allocation of IT assets (Wang et al. 2015). Thus, our hypothesis is stated as follows.

H4: Firm with the greater likelihood of managing IT will significantly improve their firm performance.

Mediating Role of Management of IT

Managing IT is a prospect to pinpoint the value of IT in business terms (Van Der Zee and De Jong 1999); it involves setting the direction for strategy and enterprise-wide coordination of IT capabilities (Karimi et al. 2000). We posit that managing IT helps to achieve firm performance through value propositions, a portfolio of services to deliver value, and production efficiency by Val-IT domain's standpoint. First, IT systems help to create a new value proposition in a way to meet customer needs and extend new offerings example, CRM application to fulfill better customer needs and customer insights for demand projection, personalized design, and manufacturing (Mithas et al. 2012). Second, managing IT as a portfolio of services which deliver business value from IT investments; subsequently, it can increase IT's contribution to achieve firm performance (Peppard 2003). Third, investment in IT improves the production efficiency of product quality, pricing decisions, lowering production costs, and productivity (Thatcher and Oliver 2001). Accordingly, Tallon et al. (2000) asserted that adopting IT management practices will bring closer to alignment between IT and business goals which in turn uplift firm performance. Moreover, investment in IT are not adequate by themselves to affect firm performance, rather it requires IT management to reconfigure IT assets into resources with their strategic potential and then set up them effectively throughout the firm (Wang et al. 2015). Thus, our hypothesis is stated as follows:

H5: The management of IT mediates on the relationship between the Val-IT components and firm performance.

Control Variables

The control variables are used to explain the factors except for the theoretical constructs, which could explain the variance in the dependent variable (Ravichandran et al. 2005). This study uses firm size (Wang et al. 2016b), firm age (Mao et al. 2016) and IT budget as the control variable. Firm size can have a great impact on firm performance (Wu et al. 2006). Firm age is perceived as an existence of inter-firm relationships, staying power, and the popularity of internal routines, all of which can affect current performance (Ravichandran et al. 2005).

## 4. Research Methodology and Data Analysis

### 4.1. Measurement Development

All five measurement constructs in the research model such as Value Governance (VG), Portfolio Management (PM), Investment Management (IM), Management of IT (MIT), and Firm Performance (FM) were adopted from the existing studies. For the Val-IT 2.0 components (VG, PM, and IM) limited literature exists. Hence, we keep consistent with the study of Ali et al. (2015) used a systematic approach by integrating Val-IT 2.0 framework with many existing literature support. This present study also includes some supportive literature for construct development in this research domain. In literature, firm performance has traditionally been considered accounting ratios as the sole indicator of performance. On the other hand, the firm performance is multidimensional in nature and accounting measures may be misleading because of ''their (1) inadequate handling of intangibles and (2) improper valuation of sources of competitive advantage'' (Bharadwaj et al. 1993; Morgan and Strong 2003). Therefore, we included measures such as financial returns, operational excellence, and marketing performance to best measure firm's total performance relative to its competition (Wu et al. 2015). Appendix 'A' lists all the constructs with their respective sources.

### 4.2. Sample and Data Collection Procedure

We collected the data from currently employed senior IT and business managers in Chinese firms. We started to collect the data from the October 2016 to mid of January

2017. Our sampling frame is all MBA and EMBA graduates in 2015 batch in the school of management, Huazhong University of Science and Technology conducted in several cities of China (Wuhan, Shenzhen, Suzhou, Nanjing, Jinan, and Guangzhou). The questionnaire link with invitation letter emailed to 650 respondents by randomly selecting 100 – 120 graduates from their alumni group in each city. The questionnaire was developed in a paid Chinese electronic platform (www.sojump.com). The respondent can answer only one questionnaire option was set to avoid the multiple responses from a single informant. Initially, 217 questionnaires were received, and the overall response rate was 33.9%. After eliminating records with incomplete and missing data, 176 valid records were selected as the sample for this study, representing 81.1% valid response rate. Our sample covers a wide range of industry sectors including manufacturing 34.7%, IT and technology 21%, hotel/restaurants 10.8%, trade and business 6.8%, banking/finance/insurance 6.8%, communication services 8.5%, transport/storage 6.2%, construction 3.4%, and others 1.8%. Table 1 shows the demographic profile of the sample.

*4.3.* Measurement Model Validation We used several methods to test the common method bias (CMB). Harman's one-factor test - If all variables load variance on one factor or one factor explains the majority of the variance; then there is a high level of common method variance (Podsakoff et al. 2003; Wang et al. 2016b). Our principal component factor analysis generated four factors; the highest variance explained by one factor is accounted for 46.2%, which is below the cutoff value 50% under Harman's single factor test (Chi et al. 2017; Turel et al. 2017). Due to the growing dispute on the merits of Harman's single-factor test we re-validated CMB using other approaches. First, if any high correlation (r > .90) is also the evidence for common method bias (Gaskin

2011; Lowry and Gaskin 2014). In any case, our Pearson's correlations r value reach this threshold (Table 3: r < 0.9). Second, if all VIFs resulting from a full collinearity test are equal to or lower than 3.3, the model can be considered free of common method bias (Kock 2015). In this study, all the VIF value are below 3.3 (Table 3: VIF < 3.3). Third, we added a common method factor (Armstrong and Overton 1977; Shao et al. 2016) with our PLS model and linked all the principal constructs' indicators with it and compared the substantive factor loadings with the method factor loadings. Then we calculated indicator's variances explained by the principal construct ($R_1^2$) and indicator's variances explained by the method construct ($R_2^2$). This shows that most of the substantive factor loadings are positive and significant while, most of the method factor loadings are insignificant. We then calculated the average variance explained by principal construct is 0.606 while the method is 0.027; subsequently compared the ratio between method and substantive variance; having a slight amount of method variance, we concluded that CMB is not a serious issue using this method. Using these approaches, the presence of the common method bias is an insignificant threat for our sample. To compare early with late responses, we defined the first 10% of the responses as early and the last 10% responses as for late (Armstrong and Overton 1977). Our t-test results show that there is no significant difference in sample characteristics between early and late response.

The Kaiser Meyer Olkin (KMO) value is 0.913 > 0.5, which indicates the measure of sampling adequacy for the data analysis, and the Barttlett's Test of Sphercity value 0.000 < 0.05 confirms the appropriateness of factor analysis (Peng et al. 2016).

*4.4. Data Analysis Method*

The partial least squares (PLS) structured equation modeling (SEM) technique was used as it efficiently handle small dataset and has greater statistical power (Hair Jr et al. 2016; Wang et al. 2016b). We used Smart PLS 3.0 for the data analysis. In this research, all the constructs are first order reflective based on the criteria suggested by Jarvis et al. (2003). Our analysis includes two steps. First, we assessed the measurement model for proper psychometric properties. The second step measures the structural model (Wang et al. 2016b).

Table 1: Demographic Profile of the Sample

|  | Category | N | % |
|---|---|---|---|
| Position | CEO | 3 | 1.7 |
|  | General Manager | 11 | 6.3 |
|  | Head of IT/MIS | 29 | 16.5 |
|  | Project Manager | 21 | 11.9 |
|  | Depart. Manager | 60 | 34.1 |
|  | Market. Manager | 34 | 19.3 |
|  | Other Managers | 18 | 10.2 |
| Experience | < 3 years | 42 | 23.9 |
|  | 3.1– 6 years | 66 | 37.5 |
|  | 6.1–9 years | 31 | 17.6 |
|  | 9.1 - 12 years | 26 | 14.8 |
|  | 12.1 - 15 years | 10 | 5.7 |
|  | 15.1 - 18 years | 1 | .6 |
| IT_budget_anual_sales | < 1 % | 50 | 28.4 |
|  | 1.1%–2% | 32 | 18.2 |
|  | 2.1%–3% | 17 | 9.7 |
|  | 3.1%–4% | 24 | 13.6 |
|  | 4.1%–5% | 19 | 10.8 |
|  | >5% | 34 | 19.3 |
| Total_sales | < 100 million $ | 35 | 19.9 |
|  | 100 - 499 million $ | 35 | 19.9 |
|  | 500 - 999 million $ | 14 | 8.0 |
|  | 1000 -1499 million $ | 16 | 9.1 |
|  | 1500 - 1999 million $ | 6 | 3.4 |
|  | ≥ 2,000 million $ | 70 | 39.8 |
| Employees | Less than 100 | 29 | 16.5 |
| | 100 – 500 | 40 | 22.7 |
| | 500 - 1000 | 15 | 8.5 |
| | 1000–1500 | 14 | 8.0 |
| | 1500 - 2000 | 5 | 2.8 |
| | More than 2000 | 73 | 41.5 |
| Org_Age | < 4.9 Years | 23 | 13.1 |
| | 5 - 9.9 Years | 35 | 19.9 |
| | 10 - 14.9 Years | 26 | 14.8 |
| | 15 - 19.9 Years | 14 | 8.0 |
| | ≥ 20 years | 78 | 44.3 |

*4.5.* Measurement Model

We measured internal consistency and reliability, convergent validity, and discriminant validity (Hair Jr et al. 2016; Wang et al. 2016b). The Cronbach's alpha values above 0.7 satisfy the requirement for internal consistency and reliability. For convergent validity, we assessed two measures a) average variance extracted (AVE) value above 0.50 and b) composite reliability (CR) value above the thresholds of 0.70 demonstrates satisfactory convergent validity. Besides, the square roots of AVE the value greater than all other cross-correlations, confirm the sufficient discriminant validity. The factor loadings all are above 0.66, signifying good indicator reliability. Table 2 shows Cronbach's Alpha values, the average variance extracted (AVE), and composite reliability (CR) of this research. Collinearity diagnostic was conducted to check the multicollinearity issue for the constructs. The variance inflation factor (VIF) for all indicators, ranging 2.147 - 3.191 (<5) which suggested a non-critical level of multicollinearity (Hair Jr et al. 2016). Hence, for our measures, multicollinearity is not a serious issue. Table 3 shows VIF values for the constructs. The control variables used in this research did not show any significant relationship.

## 5. Results and Findings

### *5.1.* Structural model and hypothesis testing

The hypothesis testing includes two steps. First, we assessed the significance of the direct paths for all constructs; in which we used subsamples of 500 for bootstrapping to analyze the significance of the path coefficients. Second, we performed mediation analysis of indirect effects of Val-IT components on firm performance through managing IT. We used $R^2$ Value in the dependent variable to measure the explanatory power of the structural model. The structural model accounted for 58.3% of the variance in managing IT, and 39.8% of the variance in firm performance which confirms the predictive validity (Hair Jr et al. 2016). Also, we conducted blindfolding to measure $Q^2$ values to confirm predictive relevance, such as for managing IT 0.403 strong, and firm performance 0.234 moderate predictive relevance that the structural model has (Hair et al. 2017; Hair Jr et al. 2016; Wang et al. 2016b). The Val -IT 2.0 components of value governance (0.158*, p <0.047), portfolio management (0.407**, p<0.000), and investment management (0.278**, p<0.002) support the hypotheses H1, H2 and H3. Accordingly, a firm with the greater the ability of value governance, portfolio management, and investment management practices will significantly associate with the management of IT. Further, this research presents evidence for the hypothesis H4 as significant and supported (0.503**, p<0.000), thus the greater the likelihood of managing IT will positively impact on firm performance. Table 4 presents structural analysis with their results.

Table 2: Construct, factor loadings, AVE and Cronbach's Alpha

| Construct | Items | Factor Loadings | *t*-Value | Cronbach's Alpha (CA) | AVE | CR |
|---|---|---|---|---|---|---|
| Val_Gov | VG1 | 0.669 | 9.942 | 0.833 | 0.668 | 0.888 |
|  | VG2 | 0.853 | 33.417 |  |  |  |
| | VG3 | 0.851 | 26.448 | | | |
| | VG4 | 0.879 | 36.416 | | | |
| Port_Mgt | PM1 | 0.850 | 26.764 | 0.852 | 0.692 | 0.900 |
| | PM2 | 0.834 | 24.280 | | | |
| | PM3 | 0.841 | 34.153 | | | |
| | PM4 | 0.801 | 20.853 | | | |
| Invest_Mgt | IM1 | 0.784 | 17.330 | 0.838 | 0.673 | 0.891 |
| | IM2 | 0.818 | 25.537 | | | |
| | IM3 | 0.842 | 31.124 | | | |
| | IM4 | 0.837 | 30.587 | | | |
| Mgt_IT | MIT1 | 0.849 | 28.992 | 0.916 | 0.749 | 0.937 |
| | MIT2 | 0.839 | 31.203 | | | |
| | MIT3 | 0.870 | 32.972 | | | |
| | MIT4 | 0.897 | 47.315 | | | |
| | MIT5 | 0.872 | 33.853 | | | |
| Firm_Perf | FR1 | 0.841 | 28.410 | 0.932 | 0.649 | 0.943 |
| | FR2 | 0.845 | 30.667 | | | |
| | FR3 | 0.848 | 29.954 | | | |
| | MP1 | 0.734 | 14.744 | | | |
| | MP2 | 0.740 | 16.932 | | | |
| | MP3 | 0.815 | 24.965 | | | |
| | OE1 | 0.810 | 28.214 | | | |
| | OE2 | 0.792 | 17.709 | | | |
| | OE3 | 0.815 | 23.951 | | | |

Table 3: Correlation, square root of AVE and VIF

| | Mean | SD | Firm_Perf | Invest_Mgt | Mgt_IT | Port_Mgt | Val_Gov |
|---|---|---|---|---|---|---|---|
| Firm_Perf | 3.496 | 0.852 | 0.805 | 3.191 | 2.654 | 2.863 | 2.270 |
| Invest_Mgt | 3.570 | 0.926 | 0.450 | 0.820 | 2.896 | _ | _ |
| Mgt_IT | 3.553 | 0.906 | 0.607 | 0.696 | 0.866 | 2.411 | 2.147 |
| Port_Mgt | 3.527 | 0.908 | 0.470 | 0.750 | 0.717 | 0.832 | _ |
| Val_Gov | 3.503 | 1.027 | 0.475 | 0.713 | 0.617 | 0.640 | 0.817 |

Note: Diagonal elements are the square root of AVE, these should exceed the inter-construct correlations for adequate discriminant validity. Values above diagonal elements are VIF value. Off-diagonal elements are the correlations among constructs

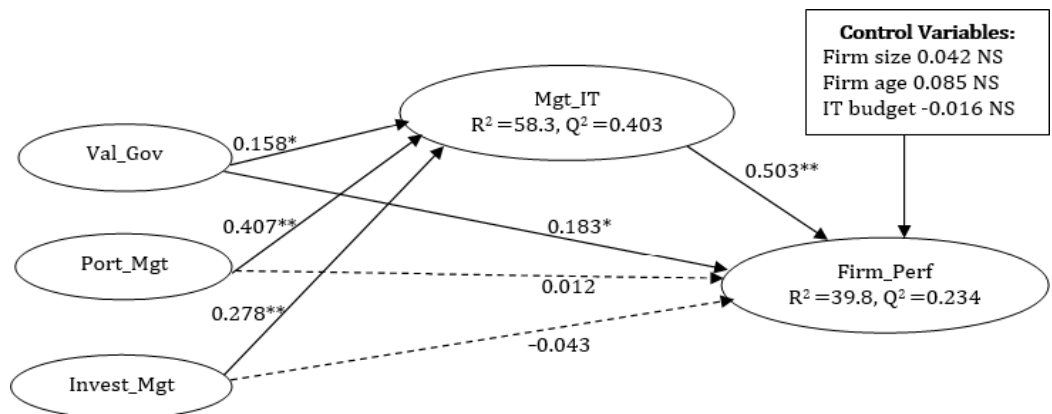

Figure 2: Path analysis results

Note: **Significance at p < 0.01. *Significance at p <0.05., NS: Not Supported

Table 4: Structural Analysis results (n=176)

|  | Direct effect | | | | | | Total effect | | |
|---|---|---|---|---|---|---|---|---|---|
| To | Management of IT | | | Firm Performance | | | Firm Performance | | |
| From | Beta | SE | p-value | Beta | SE | p-value | Beta | SE | p-value |
| Val_Gov | 0.158* | 0.073 | 0.047 | 0.183* | 0.084 | 0.024 | 0.262** | 0.088 | 0.002 |
| Port_Mgt | 0.407** | 0.095 | 0.000 | 0.012 | 0.119 | 0.919 | 0.217 | 0.118 | 0.053 |
| Invest_Mgt | 0.278** | 0.096 | 0.002 | -0.043 | 0.102 | 0.670 | 0.098 | 0.120 | 0.399 |
| Mgt_IT |  |  |  | 0.503** | 0.128 | 0.000 | 0.503** | 0.128 | 0.000 |
|  |  |  |  |  |  |  |  |  |  |
| Control Variable |  |  |  |  |  |  |  |  |  |
| Firm size |  |  |  | 0.042 | 0.068 | 0.532 | 0.042 | 0.068 | 0.532 |
| Firm age |  |  |  | 0.085 | 0.063 | 0.170 | 0.085 | 0.063 | 0.170 |
| IT budget |  |  |  | -0.016 | 0.062 | 0.777 | -0.016 | 0.062 | 0.777 |

Note: ** Significance at p < 0.01. * Significance at p <0.05.

### 5.2. Test for Mediation Effect

For mediation analysis, we followed the procedure used in the prior studies (Wamba et al. 2017; Wang et al. 2016b; Wu et al. 2015). It is based on the path coefficients and standard errors of the direct paths between (i) independent and mediating variables (i.e., iv → m), and (ii) mediating and dependent variables (i.e., m→dv)(Wamba et al. 2017).

Table 5: Test of indirect effects for mediation

|  | Total effect | | Direct effect | | Indirect effect | | | |
| --- | --- | --- | --- | --- | --- | --- | --- | --- |
|  | Beta | p-value | Beta | p-value | Beta | p-value | 95% LL | 95% UL |
| Val_Gov | 0.262** | 0.002 | 0.183* | 0.024 | 0.079* | 0.044 | 0.009 | 0.150 |
| Port_Mgt | 0.217 | 0.053 | 0.012 | 0.919 | 0.205** | 0.002 | 0.066 | 0.344 |
| Invest_Mgt | 0.098 | 0.399 | -0.043 | 0.670 | 0.140* | 0.015 | 0.016 | 0.263 |

Note: ** Significance at p < 0.01. * Significance at p <0.05.

The table 5 shows that the indirect effects are greater than direct effect except for the value governance β (0.183> 0.079) with firm performance. The portfolio management and investment management of Val-IT 2.0 has higher indirect effect such as β = 0.205, β = 0.140 respectively. To measure the mediating effects, we used the bias-corrected bootstrapping of 500 subsamples to test the indirect effects. There was no zero value between the Lower Limit (LL) and Upper Limit (UL) of 95% confidence interval of this test. The Val-IT components have positive effects value for value governance β = 0.079*, between 0.009 - 0.150, portfolio management β = 0.205**, between 0.066 - 0.344 and investment management β =0.140*, between 0.016 - 0.263 respectively. In addition, we performed Sobel test to revalidate the mediating effect (Peng et al. 2016; Sobel 1982; Wu et al. 2015). The indirect effect of Val-IT 2.0 components on firm performance are significant as indicated by the Sobel test's Z-statistics for the paths Val_Gov→Mgt_IT→Firm_Perf (Z = 1.896, p=0.0579), Port_Mgt→Mgt_IT→Firm_Perf (Z=2.896, P<0.01), and Invest_Mgt→Mgt_IT→Firm_Perf (Z=2.331, p<0.05). Therefore, our hypothesis H5 is supported that the managing IT mediates the relationship between Val-IT component and firm performance.

In this research as presumed, that the mediating role of managing IT between Val-IT components and firm performance is significant.

To determine whether managing IT completely or partially mediates the above relationship, the direct effect was examined by removing the mediator from the model. The mediation analysis implies that managing IT greatly mediates the link between Val-IT components and firm performance. The variance accounted for (VAF) determines the size of the indirect effect in relation to the total effect. It means the extent to which the variance of the dependent variable is explained by the independent variable and how much variance is explained by the indirect link through the mediator (Hair Jr et al. 2016). Accordingly, the VAF calculated values for value governance VAF = 30% - *partial mediation* and for portfolio management and investment management VAF > 80% confirm that these two have *full mediation*.

## 6. Discussion and Implications

### *6.1. Discussion*

Drawing upon IT capability and RBV, we developed a hypothetical model to examine the relationship between Val-IT components and the mediating role of managing IT on firm performance. The results show that the Val-IT components' posited relationship with the management of IT is significantly supported. It is asserted in the prior study that there is evidence for the strong link between the implementation of Val-IT and the realization of IT goals and between the achievement of IT goals and business goals (Lombardi et al. 2016). Similarly, this study finding brings potential implications that Val-IT components are aligned with the management of IT which facilitates to improve firm performance.

This research sheds further light on the significant linkage of managing IT with firm performance owing to its indispensable nature. This keeps consistent with the study of Tallon et al. (2000) that firms with more focus for IT must make greater use of certain key IT management practices in a way that adds to greater IT payoffs. Further, the study of Wang et al. (2015) shows in China that the strong IT management can directly improve firm performance. Though, IT investment constantly increases in China, it purely alone does not improve firm performance. According to Peng et al. (2016), the operational performance and competitiveness of many Chinese companies have not improved even though the multi-billion dollar has been invested in IT systems in recent years. Consequently, managing IT improves firm performance through cost reduction, quality improvement, product development, timely delivery, and higher dependability (Peng et al. 2016; Xu et al. 2016).

Further, this study shows the direct relationship of Val-IT components with firm performance in which value governance ($\beta = 0.183$, $p = 0.024$) has a significant relationship, but portfolio management ($\beta = 0.012$, $p > 0.05$) and investment management ($\beta = -0.043$, $p > 0.05$) are not significant. In the mediation analysis, the value governance is the partial mediator, but portfolio management and investment management are the full mediators. In consistent with the study of Ali et al. (2015) they modified the actual Val-IT 2.0 and linked with the corporate performance and confirmed as significant and positive. Hence, this research intends that the proper management of IT is a mediator between Val-IT domains and firm performance and it required to generate business value from IT investment. This research's theoretical and practical implications are discussed in the below sections.

*6.2. Theoretical Implication*

This study contributes to the theoretical bases of management of IT and firm performance literature in three ways. First, this research inserts into the body of Chinese IS research by discussing empirical driven firm's managing IT practice which is identified paucity in the literature. As highlighted in the opening of this paper, China is developing noticeably, hence amplified attention is being paid to the growth of IS research that will strengthen nationwide development across all sectors of the economy. There remains many fascinating IS phenomenon, e.g., IT innovation, big data, cloud computing, social computing...etc. While we believe that the existing literature gives only the tip of the iceberg for managing IT research in China. Thus, this study is the novel opening for the range of IS sensation in China. Second, we proposed and empirically corroborated the Val -IT 2.0 domains as the construct that linked through the management of IT as a mediator with firm performance. The modeling and inclusion of Val IT components in the research are unique to the literature. Hence, this study extends the literature that how Val IT 2.0 components can be integrated as a construct in the actual research. Third, this study is underlying on the theoretical facets of resource-based view and IT capabilities which are generally highlighted in the literature to demonstrate IT business value for firm performance. Although RBV and IT capabilities call for the need to study any effect on the transitional impact of IT investment on firm performance. Researchers argued that either IT investment or IT assets do not alone directly improve firm performance instead it should be combined with other capabilities, resource, management practices, and some operational strategies (Aral and Weill 2007; Peng et al. 2016). Hence, this study adopted a new approach by including managing IT as a construct to demonstrate their intermediate relationship.

*6.3. Practical Implication*

The study findings also bring certain practical implications. First, the data collected for this study is from senior level IT and business executives in China. The inclusion of corporate level people (CEO and CIO), IT, and line managers in managing IT process facilitate to share domain knowledge, IS planning and designing, IT project management and planning for IT standards and controls which in turn create synergetic decision making (Wang et al. 2015). In contrast, research shows in China most organizations have not created a position for CIOs, and instead departmental director functioning as the CIOs. As a result, they were unable to bring strategic applications, and management of IT to meet business requirements (Shao et al. 2016). Hence, this study findings are the novel direction for both formal and informal CIOs. Second, the "IT productivity paradox" in China might explain why the active performance and competitiveness of many Chinese firms have not improved hence multi-billion dollar invested in IT during recent years (Peng et al. 2016). To survive and stay competitive in the growing business world, it is vital to keep an eye on the internal performance and growths in finance, customers, market, strategies, R&D, innovation with superior management talent which nurture superior value to the firm. Thus, this study findings are deemed as a practical elucidation to Chinese IT managers and executives who suffer in deficiencies which hinder their firm performance. Last but not least, According to Peng et al. (2016), the Chinese firm's specific findings may well give a basis for developing country's firms to design and implement suitable strategies to maximize return on IT investment. Hence, this research design and findings are essential for developing, and emerging country like China, which can be generalizable to other firms in Asia, Middle East, Africa, and Europe.

## 7. Conclusion and Limitations
### 6.1. Limitations
This research has some limitations as well. First, the data were merely collected in China, which may hinder the generalization of the findings compared to other world countries' context. Second, the sample size also less as it is perceived as a limitation in other studies (Chen 2010; Kim et al. 2009; Prasad et al. 2010; Shao et al. 2016).Third, the external and environmental factors like economic, cultural, regulatory, and political and industry factors also have significant influence. The exclusion of these factors may limit this study's results. The future research avenue may also consider these factors and include mediators such as IT capabilities, IT governance mechanisms, …etc.

## 1.1. Conclusion

The rising significance of managing IT has gained extensive attention in the business world. This study adopted a broader conceptualization of RBV and IT capability to demonstrate the effect of Val IT components and mediating role of managing IT on firm performance. Finding revealed particularly, managing IT is observed to be a significant mediator which facilitates for Val-IT components effect on firm performance. The positive and significant relationship demonstrated among the constructs, and all the hypothesis are accepted. This research fills the gap in the literature that Val-IT 2.0 components rarely used in IT management empirical research, and this research offers insight facts from Chinese firm's context. This study's resource-centric approach in managing IT would give clear track for the deeper explorations which are crucial to drive firm performance.

# Appendix A

1. Construct Development Methodology

How do you rate the followings in the context of Val IT 2.0 to manage IT investment in your organization?

1. Strongly Disagree    2. Disagree    3. Neither agree or disagree    4. Agree    5. Strongly Agree

| Construct | Item Code | Measures | Source |
|---|---|---|---|
| Value Governance | VG1 | We evaluate IT investments against consistent and relevant criteria. | (Ali et al. 2015) |
| | VG2 | We review & track the benefits and costs of spending of IT investments proposals. | (Ali et al. 2015) |
| | VG3 | Has a steering committee to oversee major IT investments. | (Ali et al. 2015) |
| | VG4 | Has different stakeholder groups involved in the IT investments evaluation. | (Ali et al. 2015) |
| Investment Management | IM1 | Evaluating IT investments against a consistent and relevant set of management goals and business criteria. | (Ali et al. 2015) |
| | IM2 | Involving a steering committee (e.g. IT investments committee/board) to oversee major IT investments. | (Ali et al. 2015) |
| | IM3 | Involving different stakeholder groups (e.g., management and end-user) in the IT investments evaluation process. | (Ali et al. 2015) |
| | IM4 | Identifying the full costs associated with IT investment projects (e.g., tangible and intangible costs) and performing formal reviews after IT investments' implementations. | (Ali et al. 2015) |
| Portfolio Management | PM1 | Using sensitivity analysis (e.g., what-if analysis) for dealing with uncertainty in evaluating IT investments. | (Ali et al. 2015; Kumar et al. 2008) |
| | PM2 | Balancing the IT investments portfolio for alignment and risk-return profile. | (Ali et al. 2015) |
| | PM3 | Developing comprehensive project management metrics and regular review (e.g., costs, benefits, outcomes) for IT investments. | (Ali et al. 2015; Kumar et al. 2008) |
| | PM4 | Asking the end-users to verify that the new system meets the requirements, at the completion of the IT project. | (Ali et al. 2015; Kumar et al. 2008) |

How do you rate the following in the context of managing of IT

1. Strongly Disagree    2. Disagree    3. Neither agree or disagree    4. Agree    5. Strongly Agree

| | | | |
|---|---|---|---|
| Management of IT | MITI1 | We have set of mechanisms to request, prioritize, fund, monitor, and implement IT investment decisions to ensure IT investments deliver value to the organization. | (Wilkin et al. 2016) |
| | MITI2 | Executives and board of directors have Responsibility to ensure the organization's IT systems sustain and extend its strategies and objectives. | (Wilkin et al. 2016) |
| | MITI3 | Our company has established formal processes to govern and manage IT projects. | (Chen et al. 2014; Wu et al. 2015) |
| | MITI4 | Our company has established a formal prioritization process for IT investments and projects in which business and IT is involved. | (Wu et al. 2015) |
| | MITI5 | Our company has a Steering Committee composed of business and IT people focusing on prioritizing and managing IT projects. | (Prasad et al. 2010; Wu et al. 2015) |

How do you rate the following in the context firm performance in your organization?
1. Strongly Disagree    2. Disagree    3. Neither agree or disagree    4. Agree    5. Strongly Agree

| | | Firm Performance | |
|---|---|---|---|
| Financial Returns | FR 1 | Our company's return on investment (ROI) is better compared to other companies in the same industry. | (Prasad et al. 2010; Wu et al. 2006; Wu et al. 2015) |
| | FR 2 | Our company's return on equity (ROE) is better compared to other companies in the same industry. | (Prasad et al. 2010; Wu et al. 2015) |
| | FR 3 | Our company's return on asset (ROA) is better compared to other companies in the same industry. | (Wu et al. 2015) |
| Operational Excellence | OE 1 | Our company has better productivity improvements compared to other companies in the same industry. | (Ravichandran et al. 2005; Wu et al. 2015) |
| | OE 2 | Our company has a better timeline of customer service compared to other companies in the same industry. | (Wu et al. 2015) |
| | OE 3 | Our company has better production cycle time compared to other companies in the | (Ravichandran et al. |

|  |  |  | same industry. | 2005; Wu et al. 2015) |
|---|---|---|---|---|
| Marketing Performance | MP 1 | | Our company performs much better than competitors in sales growth. | (Wu et al. 2006) |
| | MP 2 | | Our company performs much better than competitors in market share. | (Wu et al. 2006) |
| | MP 3 | | Our company performs much better than competitors in product development and market development. | (Wu et al. 2006) |
| Control Variables | | Organization Size | | (Mao et al. 2016; Ravichandran and Lertwongsatien 2002) |
| | | Organization Age | | (Mao et al. 2016; Ravichandran and Lertwongsatien 2002) |
| | | IT budget | | |

Appendix B: Item to construct cross loadings.

|      | Firm_Perf | Invest_Mgt | Mgt_IT | Port_Mgt | Val_Gov |
|------|-----------|------------|--------|----------|---------|
| FR1  | 0.841 | 0.293 | 0.464 | 0.355 | 0.398 |
| FR2  | 0.845 | 0.379 | 0.537 | 0.409 | 0.403 |
| FR3  | 0.848 | 0.351 | 0.497 | 0.393 | 0.405 |
| MP1  | 0.734 | 0.304 | 0.390 | 0.294 | 0.312 |
| MP2  | 0.740 | 0.387 | 0.429 | 0.417 | 0.371 |
| MP3  | 0.815 | 0.379 | 0.465 | 0.374 | 0.369 |
| OE1  | 0.810 | 0.415 | 0.576 | 0.376 | 0.409 |
| OE2  | 0.792 | 0.397 | 0.470 | 0.420 | 0.409 |
| OE3  | 0.815 | 0.353 | 0.536 | 0.359 | 0.357 |
| IM1  | 0.350 | 0.784 | 0.491 | 0.557 | 0.521 |
| IM2  | 0.318 | 0.818 | 0.549 | 0.616 | 0.623 |
| IM3  | 0.441 | 0.842 | 0.627 | 0.589 | 0.567 |
| IM4  | 0.357 | 0.837 | 0.604 | 0.697 | 0.630 |
| MIT1 | 0.515 | 0.588 | 0.849 | 0.621 | 0.537 |
| MIT2 | 0.544 | 0.608 | 0.839 | 0.614 | 0.579 |
| MIT3 | 0.550 | 0.592 | 0.870 | 0.622 | 0.533 |
| MIT4 | 0.523 | 0.620 | 0.897 | 0.658 | 0.527 |
| MIT5 | 0.491 | 0.605 | 0.872 | 0.584 | 0.491 |
| PM1  | 0.368 | 0.620 | 0.556 | 0.850 | 0.465 |
| PM2  | 0.351 | 0.623 | 0.536 | 0.834 | 0.476 |
| PM3  | 0.455 | 0.652 | 0.670 | 0.841 | 0.612 |
| PM4  | 0.374 | 0.595 | 0.604 | 0.801 | 0.555 |
| VG1  | 0.308 | 0.476 | 0.312 | 0.397 | 0.669 |
| VG2  | 0.425 | 0.614 | 0.533 | 0.581 | 0.853 |
| VG3  | 0.361 | 0.567 | 0.521 | 0.524 | 0.851 |
| VG4  | 0.443 | 0.657 | 0.599 | 0.567 | 0.879 |